\documentclass[aps,groupedaddress]{revtex4}
\usepackage{graphicx}
\newcommand{\be}{\begin{equation}}
\newcommand{\ee}{\end{equation}}
\newcommand{\ba}{\begin{eqnarray}}
\newcommand{\ea}{\end{eqnarray}}\def\ket#1{\left\vert #1 \right\rangle}

\begin{document}\title{Energy protection arguments fail in the interaction picture}
\author{Kenneth R. Brown}
\affiliation{School of Chemistry and Biochemistry and Computational Science and Engineering Division,Georgia Institute of Technology, Atlanta, Georgia 30332, USA}

\date{\today}
\begin{abstract}

Spin Hamiltonians with degenerate ground states are one potential system  for the storage of quantum information at low temperatures. Trapped ions can be used to simulate the dynamics of these Hamiltonians, but the coherence-preserving properties will be lost. This illustrates that a quantum simulation performed in an interaction frame will not thermalize with its environment.  

\end{abstract}
\pacs{}\maketitle

\section{Introduction}

The preservation of quantum information is difficult. The system must be protected from noise due to its environment and from errors induced by control and measurement. In comparison, classical information is robust. Usually, there is a large energetic/kinetic barrier separating the two states relative to the temperature. A common example is the use of magnetic domains for computer memory \cite{Ross:99}. 


Motivated by the idea of magnetic domains and topological particles, a number of ground state degenerate quantum systems have been proposed as error free memories \cite{Kitaev:97b,Barnes:00,supercoherent,Wang:03,Ioffe:02, Bacon:06}. In each case, the logical qubit is the degenerate ground state of many physical qubits in the presence of a Hamiltonian, $H$. The environmental noise interacts with the qubits independently and $H$ is constructed such that all single qubit errors raise the energy. Consequently at a temperature lower than the energy gap, multi-order processes are required to decohere the system. Additionally, although the lowest order processes do not preserve the exact state, they often preserve the superposition over large subspaces. This is equivalent to quantum error correction \cite{Gottesman:97a} where the distance between encoded states allows one to detect and correct up to $n$ errors. The ground states of these Hamiltonians can be used to derive quantum codes for quantum error correction \cite{Bacon:06, Cross:07}. However, the goal is to avoid active error correction and let the energetics maintain the coherence. In these systems, the error rate is suppressed as $1/\Delta^k$ where $\Delta$ is the energy gap between the ground state and $k$ depends on the size of the system and the details of the Hamiltonian.

The Hamiltonians used for protecting coherences are traditionally associated with condensed matter physics but have now been proposed for atomic and molecular physics using arrays of ions, atoms, and molecules\cite{Porras:04, Micheli:06, Rey:07}. Atomic ions have already been shown to have long lived qubits relative to the frequencies of these effective Hamiltonians \cite{Langer:05}. Is it possible that these methods can be used to extend these lifetimes even longer?

A problem arises because the effective Hamiltonians are in a rotating frame relative to the Bohr frequency of the physical qubit. Although the energy gap may protect the system against noise of frequency lower than $\Delta$, it will not suppress the natural spontaneous emission. The utility of such a scheme then depends strongly on the kinetics.

In this paper, the effects of spontaneous emission in these systems is examined by looking at a model two qubit system.  We find that the interaction picture is equivalent to a system in an environment with negative energy modes. We then examine the implications for energy-preserving Hamiltonians, quantum simulations, and adiabatic quantum computations in the interaction picture.

\section{2 qubits coupled by $XX+YY$}

\subsection{Model System}

The model systems is described by the two qubit Hamiltonian

\be
H_{xy}=\frac{\hbar J}{2}(X_1X_2+Y_1Y_2)
\ee

where $X$ and $Y$ are Pauli spin operators. The eigenstates of the Hamiltonian are $\ket{\psi_1}=\frac{1}{\sqrt{2}}(\ket{01}-\ket{10}), \ket{\psi_2}=\ket{00},\ket{\psi_3}=\ket{11},$ and $\ket{\psi_4}=\frac{1}{\sqrt{2}}(\ket{01}+\ket{10})$ with eigenenergies $E_1=-\hbar J, E_2=E_3=0,$ and $E_4=\hbar J$.

Although there is no protected qubit since the ground state is not degenerate, $\ket{\psi_1}$ has the property that any single qubit Pauli error on $\ket{\psi_1}$ raises the energy by at least $\hbar J$. Therefore, we expect that the state $\ket{\psi_1}$ will be preserved for temperatures lower than $\hbar J/k_b$. We can quantify this intuition by assuming each qubit interacts with an independent bath of harmonic oscillators and calculating the transition rate from $\ket{\psi_1}$ to all other states.

The Hamiltonian of the system and the environment as 

\be
H=H_{xy}+\sum_{i,k} \hbar\omega_k a_{ik}^\dagger a_{ik}+\hbar\sum_i \alpha_{ikz}Z_i(a^\dagger_{ik}+a_{ik})+\alpha_{ikx}X_i(a^\dagger_{ik}+a_{ik})+\alpha_{iky}Y_i(a^\dagger_{ik}+a_{ik})
\ee

where $\alpha_{iki}$ are the interaction strengths of mode $k$ with qubit $i$ along the axis $j$.

We can transform to an interaction picture with respect to the bath Hamiltonian, $H_0=H_b=\sum_{i,k}\hbar\omega_k a_{ik}^\dagger a_{ik}$. The Hamiltonian in the interaction picture in general is $\tilde{H}=\exp(iH_0t/\hbar)(H-H_0)\exp(-iH_0t/\hbar)$. For this case, we find
\be
\tilde{H}=H_{xy}+\hbar\sum_{i,k} \alpha_{ikz}Z_i(a^\dagger_{ik}e^{i\omega_kt}+a_{ik}e^{-i\omega_kt})+\alpha_{ikx}X_i(a^\dagger_{ik}e^{i\omega_kt}+a_{ik}e^{-i\omega_kt})+\alpha_{iky}Y_i(a^\dagger_{ik}e^{i\omega_kt}+a_{ik}e^{-i\omega_kt})
\ee

Fermi's golden rule is used to calculate the first order transition rate from the ground state to the other states assuming a generic spectral density $\rho(\omega)$ and thermal occupation, $n(\omega,T)=\left[e^{\frac{\hbar\omega}{k_bT}}-1\right]^{-1}$. We find that
\ba
\Gamma=4\pi\alpha_z(2J)^2\rho(2J)n(2J,T)+4\pi(\alpha_x(J)^2+\alpha_y(J)^2)\rho(J)n(J,T)
\ea
where $\alpha(\omega)$ is the interaction strength averaged over the modes of frequency $\omega$. This is the heart of these protection schemes. Because we are in the ground state, energy must be absorbed. At low temperatures, a linear decrease in the temperature  will lead to an exponential decrease in $n(J,T)$ and the transition rate.  At zero temperature $n(J,T)$=0 and the rate of excitation from the ground state vanishes to first order. In this case, because the ground state is non-degenerate, the higher-order rates will also be zero. 

\subsection{Effective Hamiltonian with Trapped Ions}

An effective $XX$ or $YY$ coupling between two ions has been demonstrated by virtually exciting the shared motional mode \cite{Molmer:98, Haljan:05L}. Although, the same motional mode cannot be used to generate an $XX+YY$ coupling, it is possible to use the center of mass mode of the two ions for the $XX$ coupling and the stretch mode for the $YY$ coupling. This yields again the Hamiltonian, $H_{xy}$. The coupling strength $J/2=\eta^2\Omega^2/\delta$ where $\eta$ is the Lamb-Dicke parameter, $\Omega$ is the Rabi frequency of the qubit transition and $\delta$ is the detuning of the laser from the relevant motional sideband. $J$ is limited to be smaller than the ion trap motional frequency $\omega_s$. In practice, $\omega_s/2\pi$ is typically a few MHz. 


Implicit in the above description is that $H$ is in the rotating frame or interaction picture defined by the Bohr frequency of the physical qubit, $\nu$. For concreteness, consider the optical qubit of $^{40}$Ca$^+$ where the $D_{5/2}$ state is $\ket{0}$ and the $S_{1/2}$ state is $\ket{1}$ \cite{SchmidtKaler:03}. For this quadrupole transition,  $\nu/2\pi=$ 411 THz and $\gamma/{2\pi}= 0.16$ Hz. 

\begin{figure}
\includegraphics{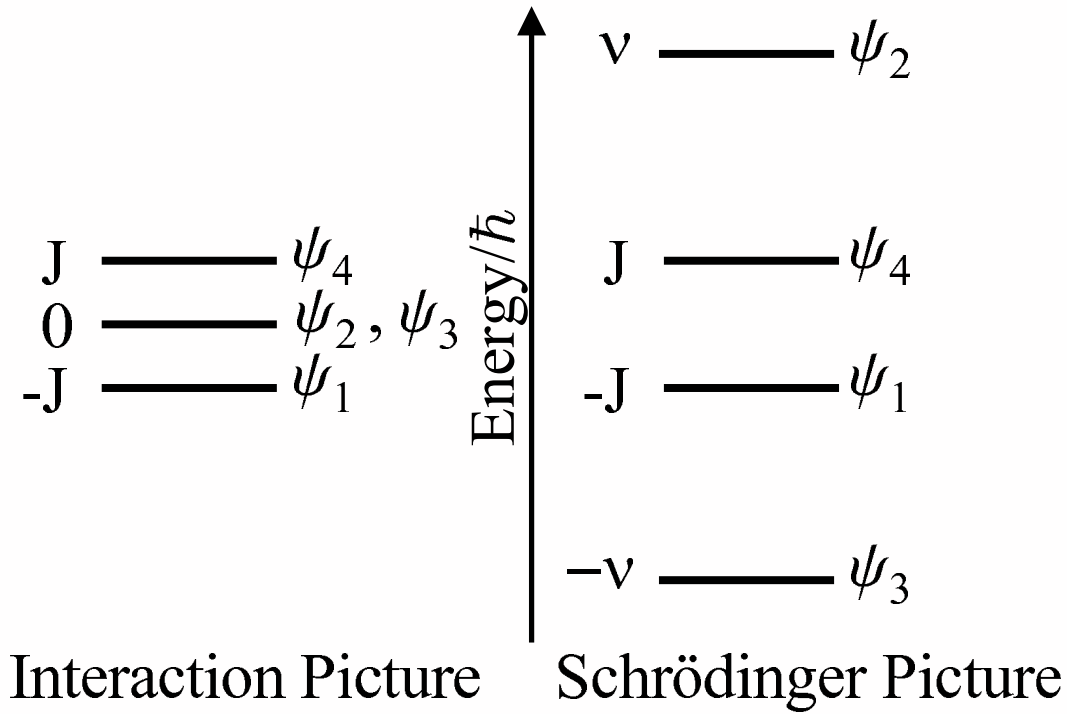}
\caption{Comparison of the effective system spectra in the interaction picture and Schr\"{o}dinger picture. The ground state of the effective Hamiltonian $\ket{\psi_1}$ is not the ground state of the total system, $\ket{3}$. The figure is not to scale. For optical qubits, $\nu/J \approx 10^8$, and for hyperfine qubits, $\nu/J \approx 10^3$.}
\label{twopics}
\end{figure}

The state $\ket{\phi_1}$ will not be preserved. Spontaneous emission will drive the system towards $\ket{\phi_2}=\ket{11}$. Although, $\ket{\phi_1}$ is the ground state of $H_{xy}$, it is an excited state of the system Hamiltonian $H_s=\frac{\hbar\nu}{2}Z$, Figure \ref{twopics}. We start with the same system-bath interaction as above but we then transform to an interaction picture defined by the uncoupled system and bath Hamiltonians $H_0=H_s+H_b$. The transformed  Hamiltonian is then

\be
\tilde{H}=H_{xy}+\hbar\sum_{i,k} \left(\alpha_{ikz}Z_i a^\dagger_{ik}e^{i\omega_kt}+(\alpha_{ikx}+i\alpha_{iky})\sigma_{-i}e^{-i\nu t}(a^\dagger_{ik}e^{i\omega_kt}+a_ke^{-i\omega_kt})+ H.c. \right)
\ee
where $\sigma_{-i}=(X_i+iY_i)/2$ is the raising operator on the $i$th qubit and $H.c.$ is the Hermitian conjugate. 

\begin{figure}
\includegraphics{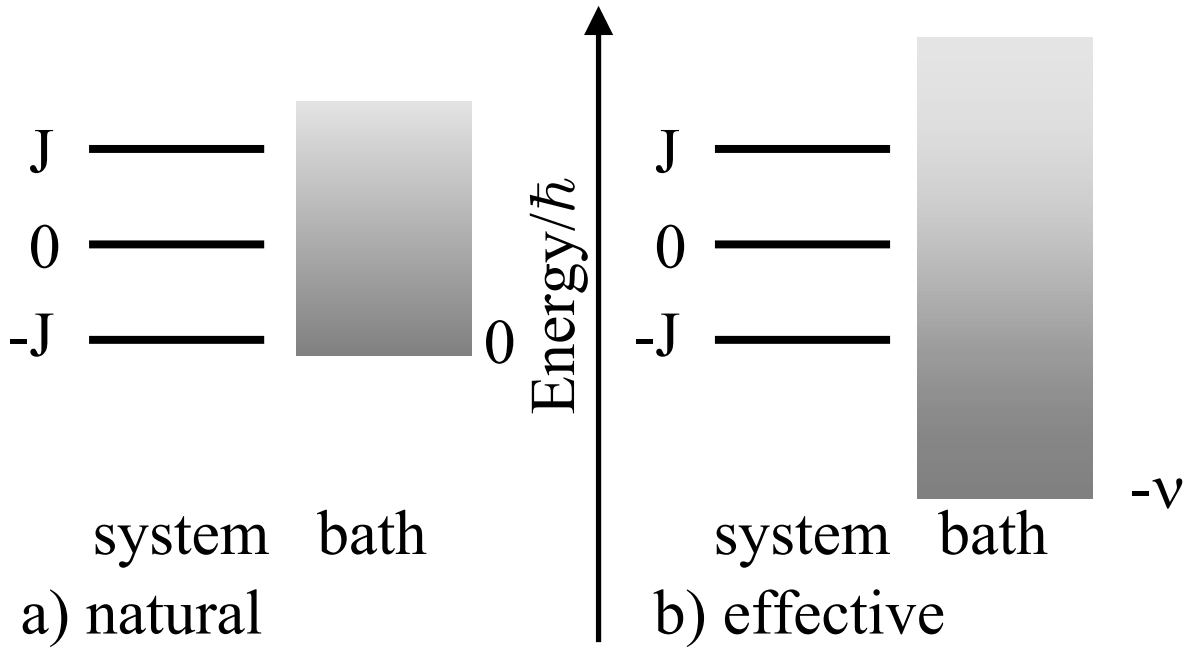}
\caption{Comparison of the system and bath spectra. We assume a physical bath of positive energy modes. a) The expected spectrum for two qubits naturally coupled by $H_{xy}$. If the temperature is less than the energy gap to the excited state, excitations will be exponentially suppressed. b) In an interaction frame, where the effective interaction is $H_{xy}$ the system spectrum is the same. However, the bath modes are shifted to effective negative energy. Consequently, a negative energy Boson can be emitted allowing for the spontaneous excitation of the system.}
\label{neg}
\end{figure}

Notice that in the interaction picture, the bath frequencies have been effectively shifted by a frequency $\pm \nu$, Figure \ref{neg}. This is equivalent to the bath having modes of negative energy. In this situation, spontaneous emission can lead to heating of the system. We again calculate the transition rates
 
\ba\nonumber
\Gamma&=&4\pi\alpha_z(2J)^2\rho(2J)n(2J,T)+2\pi(\alpha_x(\nu+J)^2+\alpha_y(\nu+J)^2)\rho(\nu+J)n(\nu+J,T)\\ \nonumber &+&2\pi(\alpha_x(\nu-J)^2+\alpha_y(\nu-J)^2)\rho(\nu-J)(n(\nu-J,T)+1)\\
\ea
At zero temperature, we have a non-zero transition rate
\ba
\Gamma&=&2\pi(\alpha_x(\nu-J)^2+\alpha_y(\nu-J)^2)\rho(\nu-J).
\ea
and as expected $\Gamma\approx\gamma$ when $J$ is small compared to $\nu$, as is the case in ion traps. We see that spontaneous emission drives the system out of the ground state of the effective Hamiltonian and the state has a lifetime equivalent to the excited state of a single ion. 

Theoretically, we can examine the case where the ions emit into the same photon mode. In this case, photon emission from the state $\ket{\psi_1}$ is prevented by destructive interference, $\gamma_{sub}=0$. This is perfect Dicke subradiance \cite{Dicke:54}  and an example of a 1-dimensional decoherence-free subspace \cite{Zanardi:97c,Lidar:98}. In this situation, the applied Hamiltonian would reduce the errors associated with low frequency errors.  In practice, ions are a few $\mu$m apart from each other and subradiance is a weak effect, $\gamma_{sub}=0.99 \gamma$ \cite{Devoe:96}.

For optical ion qubits, the applied Hamiltonian does not reduce the errors. Quantum information ion trap experiments often use hyperfine states instead of optical qubits \cite{Wineland:98}. A hyperfine qubit is a naturally energetically protected subspace with respect to electronic transitions. Spontaneous emission via a magnetic transition is possible but is negligible on the time scale of an experiment. The main source of environmental decoherence in an ion trap experiment is uncontrolled magnetic fields \cite{Berkeland:98}. This decoherence can be minimized by encoding into clock states \cite{Haljan:05A,Langer:05}. Applying the Raman beams necessary to generate $H_{xy}$ introduces a new error pathway due to Raman scattering \cite{Ozeri:05}. Again, this error is not suppressed by the applied $H_{xy}$ because the energy shift associated with $J$ is small compared to the frequency of the scattered photons.

\section{Implications of spontaneous emission for the interaction picture}

\subsection{No energy protection}\label{noprot}

A long-range quantum compass model has been proposed to protect qubits in an ion trap \cite{Milman:07}. Contrary to Ref. \cite{Milman:07}, spontaneous emission is not suppressed even though the emission occurs at a slow enough rate such that eigenstates of the interaction Hamiltonian will decay to eigenstates of the interaction Hamiltonian. Here we examine the effects of spontaneous emission for the 4 qubit system. In this case, the system is equivalent to the compass model \cite{Dorier:05} and the Hamiltonian is

\be
H=\hbar J(X_1X_3+X_2X_4+Y_1Y_2+Y_3Y_4)
\ee 

The Hamiltonian commutes with $X_1X_2$ and $X_3X_4$. The eigenvalues of these operators label four four-dimensional subspaces, $A_{ij}$. The logical $\ket{0}$ and $\ket{1}$ are the ground states of the subspaces $A_{++}$ and $A_{--}$, respectively. It could be possible that spontaneous emission, although exciting the ground state, may preserve the superpositions between the subspaces. Unfortunately, spontaneous emission connects the subspace $A_{\pm\pm}$ to both subspaces $A_{+-}$ and $A_{-+}$.

The eigenstates, $\ket{\psi_j}$, for all the subspaces are computed and the rates between the states are calculated to be proportional to $(\nu-(E_j-E_i)/\hbar)^5 |q_{ji}|^2$ where $(q_{ji})$ is the sum of the individual quadrapole moments evaluated for the eigenstates $i$ and $j$. We note that $(E_j-E_i)/\hbar$ is on the order of $J$ and in ion traps always much less that $\nu$ (even for hyperfine qubits). Therefore, in our calculations we approximate $(\nu-(E_j-E_i)/\hbar)^5\approx \nu^5$.

\begin{figure}
\includegraphics{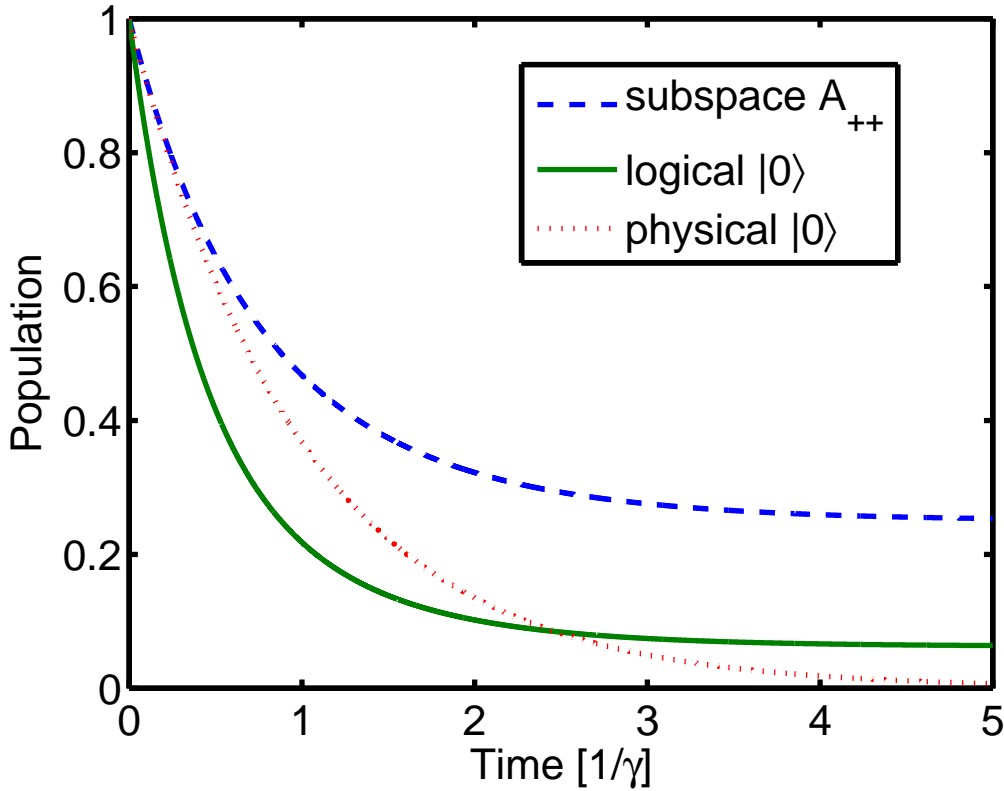}
\caption{(Color online) The compass model in the interaction picture does not suppress errors due to spontaneous emission. The decay of the population of the logical $\ket{0}$ (solid line) is comparable to the decay of the excited state in the physical qubit (dotted line). The same is true if we consider the population of the subspace $A_{++}$ associated with the logical $\ket{0}$. The applied compass model Hamiltonian drives the system from the bare ground state (all ions in the $\ket{S}$ state) and the result to zeroth order in $J/\nu$ is a completely mixed state.}
\label{decay}
\end{figure}

The population dynamics are calculated starting from the logical state $\ket{0}$. The population randomly walks among the eigenstates with a rate of the natural spontaneous emission. The state that does not spontaneously emit, $\ket{1111}$, is not an eigenstate of the effective Hamiltonian; the resulting steady state is a dynamic equilibrium with equal population in all states. Figure \ref{decay} shows the result of this calculation. The two measures used are the population in the state $\ket{0}$ and the extent of the population in the subspace $A_{++}$. We see that in a time $5/\gamma$ (5 s for $^{40}$Ca$^{+}$) the populations are almost completely mixed. Compared to the spontaneous emission rate from the $\ket{0}$ state of a single qubit there is no improvement.

\subsection{Quantum simulations improperly thermalize}

A quantum simulation uses a controlled quantum system to model another quantum system \cite{Lloyd:96}. The principle of these devices have been demonstrated in the interaction picture using nuclear magnetic resonance \cite{Cory:00,Laflamme:04,bcsnmr}. Trapped ions are seen as a promising candidate to simulate various magnetic systems \cite{Porras:04} including the compass model described above. Unlike state preservation, a quantum simulation could benefit from noise. Classical techniques for describing noisy quantum systems are often computationally more expensive. For example, calculating the dynamics of a system under the Lindblad equation \cite{Lindblad:76} requires quadratically more resources when compared to the Schr\"{o}dinger equation. Although in the limit of large noise a quantum system can be efficiently classically simulated \cite{Aharonov:96a}, there should be a range of noisy quantum simulations that are still more efficient than a classical simulation.

In the interaction picture the effective noise model is often not physical. 
In Section \ref{noprot}, we calculated that the dynamic equilibrium of the compass model with spontaneous emission was the completely mixed state. Although the physical bath is at zero temperature, the effective temperature of the system is infinite. For the two qubit example, the final state of the system is non-Boltzmann distributed with respect to the effective Hamiltonian. Ion trap and nuclear magnetic resonance quantum simulations require that the simulation be over before the state decoheres.

Quantum simulations often employ an initialization by adiabatic evolution \cite{Porras:04,bcsnmr}. Adiabatic quantum computation \cite{Farhi:00} can be viewed as a special case of a quantum simulation where the ground state of the simulated Hamiltonian solves a computational problem. The ability of the system to follow the ground state is determined by the speed of the evolution and the minimal gap between the ground and first excited state. Although a noisy environment can ruin the adiabatic algorithm when the gap becomes small \cite{Sarandy:05}, the thermalization of the system may erase these errors by recooling to the ground state as the gap becomes large \cite{Childs:01}. Unfortunately, in the interaction picture a thermal state is not achieved. Instead, spontaneous emission drives the simulation / adiabatic quantum computation to a dynamic equilibrium that may not be the desired state.

\section{Conclusions}

The simulation of a system with Hamiltonian, $H_c$, is the mapping of the dynamics of $H_c$ onto another system. Experience with how the inital system interacts with a physical environment leads us to assume that the simulated dynamics will decohere in a similar way. This intuition is incorrect. Often the simulated dynamics are in a rotating frame or interaction picture. In this frame, the bath modes appear to have negative energy and the ground state of $H_c$ is not a steady state even at zero temperature.

Trapped atomic ions have been used to demonstrate the error-reducing techniques of quantum error correction \cite{Chiaverini:04} and decoherence-free subspaces\cite{Kielpinski:01}. However, the technique of constructing a multi-body Hamiltonian with a naturally preserved ground state is incompatible with how multi-qubit gates are implemented in ion traps. As a result, the effective Hamiltonians do not yield the coherence properties associated with the natural Hamiltonians proposed in \cite{Kitaev:97b,Barnes:00,supercoherent,Wang:03,Ioffe:02, Bacon:06,Milman:07}. The realization of coherence-perserving Hamiltonians will require the construction of sophisticated condensed matter quantum information systems \cite{Ioffe:02,Weinstein:05}.

\section{Acknowledgments}
K.R.B. thanks Dave Bacon for many useful discussions.
This work was supported by the Georgia Institute of Technology.

\bibliographystyle{apsrev}

\end{document}